\documentclass{iau}

\usepackage{amsmath}
\usepackage{graphicx}
\usepackage{multirow}
\usepackage{natbib}

\usepackage{wasysym}

\begin{document}

\lefttitle{C. Norambuena, C. Charalambous and C. Aguilera-Gómez}
\righttitle{White dwarf pollution: the role of planetary mass}

\jnlPage{1}{7}
\jnlDoiYr{2026}
\doival{10.1017/xxxxx}

\aopheadtitle{Proceedings IAU Symposium}
\editors{G. Buldgen, A. Vidotto \& A. Miglio, eds.}

\title{White dwarf pollution in wide binaries: the role of planetary mass}

\author{C. Norambuena, C. Charalambous and C. Aguilera-Gómez}
\affiliation{Instituto de Astrofísica, Pontificia Universidad Católica de Chile, Av. Vicuña Mackenna 4860, 782-0436 Macul, Santiago, Chile}

\begin{abstract}

Atmospheric pollution in white dwarfs is interpreted as evidence for the accretion of rocky material from remnant planetary systems. In wide binaries, these reservoirs can be influenced by planetary perturbers. Here, we investigate how planetary mass affects the survival and delivery of planetesimals onto an evolving white dwarf.    
    
We perform N-body simulations incorporating time-dependent stellar evolution and a stellar companion, varying planetary masses and following the evolution of a planetesimal disc.

We find that lower-mass planets perturb the planetesimal reservoir less efficiently, allowing more planetesimals to survive longer. Consequently, delivery onto the white dwarf can persist into the cool white dwarf phase, whereas high-mass planets deplete the reservoir earlier, preventing sustained pollution at late stages. Our results show that planetary mass  regulates the timescale over which rocky material remains available for accretion, while the planet's location has minor impact. Pollution is more frequent in binaries with lower inclinations and eccentricities.

\end{abstract}


\maketitle

\section{Introduction}
\label{sect:intro}

More than 95\% of stars in the Galaxy are expected to end their lives as white dwarfs (WDs). These stellar remnants typically have masses of $\sim0.6\,M_\odot$ and radii comparable to that of Earth. Owing to their strong surface gravity, heavy elements rapidly sink below their thin atmospheres \citep{1995Chayer, 2009Koester}, leaving hydrogen or helium as the dominant observable constituents. However, the presence of planetary material can produce detectable atmospheric metal pollution, with roughly 25--50\% of single WDs showing evidence of such pollution \citep{2014Koester, 2024OuldRouis}.

This pollution is thought to arise from the disruption and accretion of rocky bodies from surviving planetary systems \citep{2007Zuckerman,2013Farihi,2014Jura,2014Xu}.  Planetesimals perturbed onto star-grazing orbits can be tidally disrupted and form an accretion disc around the WD, providing a means of probing the composition of extrasolar rocky material \citep{2016Farihi,2025A&A...693A..64A}. Observational evidence supports this scenario, including the detection of circumstellar debris discs through infrared excess around polluted WDs \citep[e.g.,][]{2024Rogers}, as well as observations of disintegrating planetary material transiting WDs \citep[e.g.,][]{2015Vanderburg,2020Vanderbosch,2025Guidry}. The observed diversity of polluted WDs therefore provides evidence that planetary systems can retain substantial reservoirs of minor bodies long after their host stars have evolved off the main sequence (MS).

As planetary systems and debris discs are common around MS stars, a fraction of their rocky material is expected to survive post-MS evolution and remain during the WD stage \citep{2018Mustill}. The long-term survival and dynamical evolution of these remnant systems are shaped by stellar mass loss and tidal interactions during the giant phases \citep{2016Veras}. Numerical studies have shown that stellar mass loss can destabilize remnant planetesimal belts \citep{2011Bonsor} or allow planetary systems to survive through the red giant branch (RGB) and asymptotic giant branch (AGB) phases \citep{2012Mustill,2020Ronco}. Together, these studies highlight the role of post-MS stellar evolution in shaping planetary architectures that may ultimately lead to WD pollution.

Although some polluted WDs are found in wide binary systems, most are not known to have stellar companions \citep[Planetary Enriched White Dwarf Database, PEWDD;][]{2024Williams}. The observed binary fraction depends strongly on sample selection: while some polluted-WD samples have binary fractions comparable to that of the field, fractions of only $\sim2\%$ have been reported for the most strongly polluted populations \citep{2019Wilson,2024Noor}. This suggests that the connection between binarity and pollution may depend not simply on the presence of a companion, but on the binary architecture and the properties of the polluted WD population. Wide companions can induce secular perturbations that either enhance or suppress planetesimal delivery onto the WD, depending on the binary configuration \citep{2018Veras,2026AvilaBravo}. Understanding these effects is therefore important for interpreting the occurrence and properties of polluted WDs in binary systems.

The dynamical role of wide stellar companions in WD pollution has been explored in several studies. Wide companions can induce von Zeipel-Kozai-Lidov \citep[ZKL,][]{1910vonZeipel,1962Kozai,1962Lidov}  oscillations that drive planets and planetesimals onto highly eccentric orbits, either following planetary engulfment during the AGB phase \citep{2017Petrovich} or as a consequence of post-MS stellar evolution and mass loss \citep{2017Stephan}. Direct N-body studies have further shown that stellar mass loss can destabilize initially stable planetary systems, with the resulting stability boundaries depending strongly on the binary architecture \citep{2017Veras}. 
These studies highlight the important role of wide companions in shaping the long-term dynamical evolution of planetary systems and their potential to pollute WDs.

The dynamical evolution of remnant planetary systems is shaped by both the planetary architecture and the perturbations from stellar companions. Here, we focus on the role of planetary mass, which determines the strength of the gravitational perturbations exerted on surrounding planetesimals and may therefore affect the lifetime of the planetesimal reservoir and the duration of WD pollution. We vary the mass of a planet and explore different binary eccentricities and inclinations, following the evolution of the planetesimal disc through the giant and WD phases. We investigate how planetary mass and binary architecture influence the long-term survival of the reservoir and the delivery of rocky material onto the WD. In Sec.~\ref{sect:methods} we present our simulation setup, in Sec.~\ref{sect:results} we present and analyse the results of our parameter exploration. We discuss the implications of these results and summarize our main findings in Sec.~\ref{sec:conclusion}.

\section{Methods}
\label{sect:methods}

We consider a hierarchical system consisting of a $2\,M_{\odot}$ host star orbited by a planet and a planetesimal disc, with a $1\,M_{\odot}$ stellar companion at $a_{\rm B}=250$ au. The lower-mass companion remains essentially unevolved over the timescale considered and therefore acts primarily as a gravitational perturber. We follow the dynamical evolution of the planetary system as the host star evolves from the post-RGB to the WD phase.

We combine stellar evolution models from \texttt{MESA} \citep{Paxton2011} with direct N-body integrations using \texttt{REBOUND} \citep{2012ReinLiu}. The host-star evolution is computed from the zero-age main sequence (ZAMS) to the WD phase, providing the time-dependent stellar mass, radius, and luminosity. Mass loss during the RGB and AGB phases is described by the \citet{1975Reimers} and \citet{1995Blocker} prescriptions, with $\eta_{\rm RGB}=0.35$ and $\eta_{\rm AGB}=1$, respectively. We evolve the star until it reaches a final mass of $M_{\rm WD}=0.6\,M_{\odot}$.

We distinguish between hot (WD$_{\rm h}$) and cool (WD$_{\rm c}$) white dwarfs, with $T_{\rm eff} \geq 25,000$ K and $T_{\rm eff}<25,000$ K, respectively. This threshold marks the regime in which radiative levitation can support metals in the atmosphere \citep{1995Chayer,2009Koester,2016Farihi}. We therefore consider metal pollution in WD$_{\rm c}$ to be more reliably associated with external accretion of rocky material. The simulations are continued until the WD cools to $T_{\rm eff}=5,000$ K (approximately 5\,Gyr), below which WDs become increasingly difficult to characterize spectroscopically due to their low luminosity \citep[$L<10^{-3}L_{\odot}$,][]{2017Hollands,2022Elms}. 

Stellar mass-loss and tidal interactions act in opposite directions on the orbital evolution of close-in bodies. As the host star loses mass during its giant phases, planetary orbits expand approximately adiabatically. In contrast, tidal dissipation transfers angular momentum between the stellar spin and the orbit and can drive orbital decay, particularly as the stellar radius increases during the giant phases. The competition between these processes therefore plays a key role in determining whether planets and planetesimals survive or are engulfed by the evolving star. We model tides using the weak-friction, constant time-lag formalism \citep{1979Mignard,1981Hut}, with convective damping appropriate for evolved stars \citep{1977Zahn}. The time-dependent tidal lag is computed from the stellar mass $M(t)$, radius $R(t)$, and luminosity $L(t)$ provided by \texttt{MESA}, as
\begin{equation}
    \tau=\frac{2R^3}{GMt_f},
    \qquad
    t_f=\left(\frac{MR^2}{L}\right)^{1/3}.
\end{equation}
These effects are incorporated into the N-body integrations through \texttt{REBOUNDx} \citep{2020Tamayo} and its parameter interpolation module \citep{2022Baronett}.

We explore binary inclinations $i_{\rm B}=0^\circ-90^\circ$ and eccentricities $e_{\rm B}=0-0.9$. The planet is initially placed at $a_{\rm p}=3$ or $5$~au on a quasi-circular orbit, while the planetesimal disc extends from $[1.25,2.5]$ or $[3.5,4.5]$~au, respectively. The disc contains 25 massive planetesimals, each with a mass of $4\,M_{\leftmoon}$, which gravitationally interact with one another, the planet, and the binary companion throughout the simulations.

To investigate the role of planetary mass, we vary the mass of the planet while keeping the remaining system parameters fixed. The simulations are performed using the \texttt{TRACE} hybrid integrator \citep{2024Lu}, which combines \texttt{WHFAST} \citep{2015ReinTamayo} for regular orbital evolution with the adaptive \texttt{IAS15} integrator \citep{2015ReinSpiegel} during close encounters.

\section{Results}
\label{sect:results}

\begin{figure}
    \centering
    \includegraphics[width=0.9\linewidth]{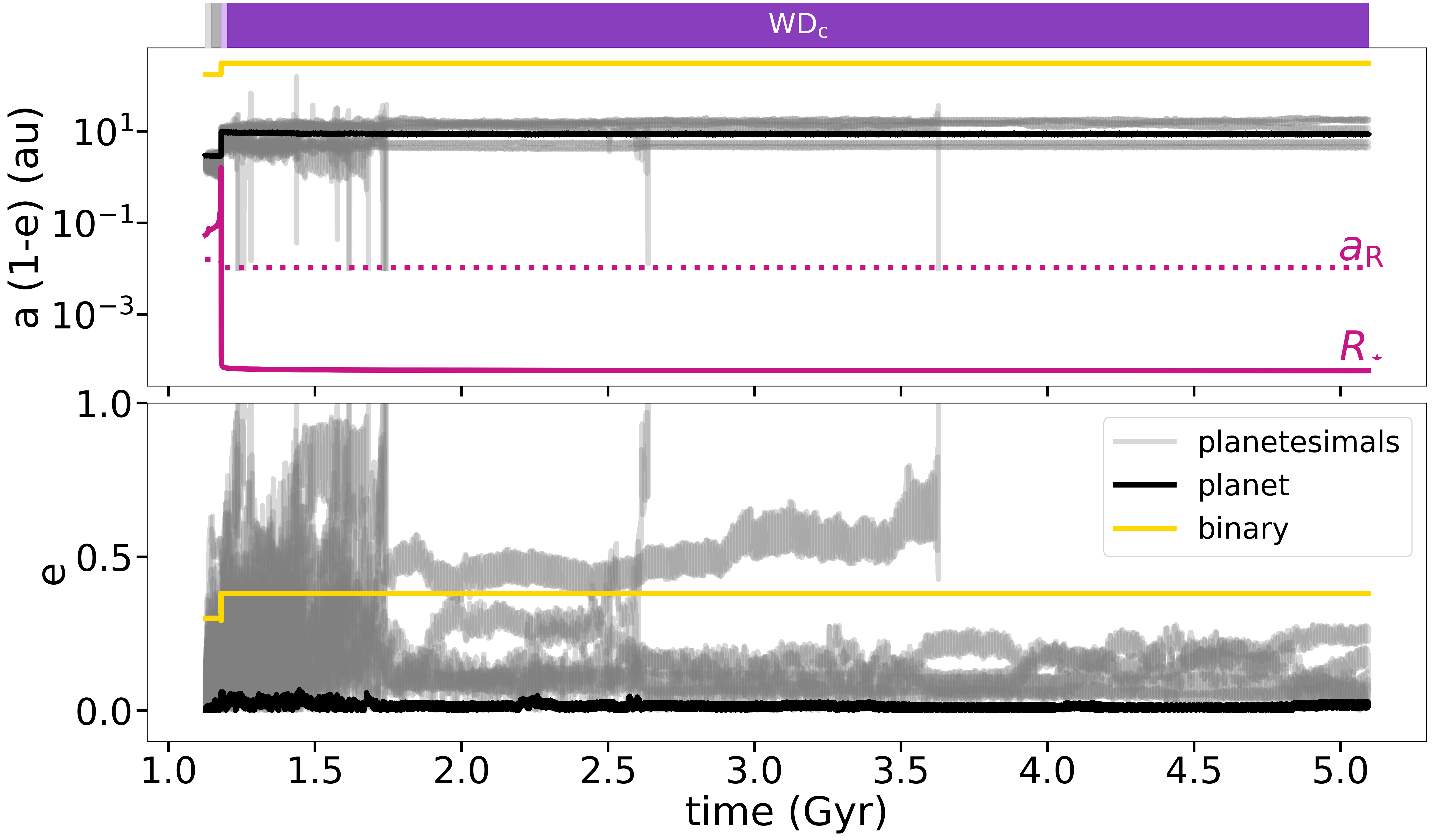}
    \caption{Temporal evolution of the pericenter (top) and eccentricity (bottom) of all bodies in a system, with time corresponding to the host star's age, counting from the ZAMS. The initial binary configuration has $a_{\rm{B}} = 250$~au, $i_{\rm{B}} = 0^\circ$, and $e_{\rm{B}} = 0.3$ (yellow). A $10\,m_\oplus$ planet (black) is located at $a_{\rm{p}} = 3$~au, while the planetesimal disc (gray) is initially placed interior to it, spanning $a \in [1.25, 2.5]$~au. The colour bar indicates the evolutionary stages of the host star. In the top panel, the pink solid and dashed lines represent the stellar radius and Roche limit, respectively.}
    \label{fig:temporal_evolution}
\end{figure}

\begin{figure}
    \centering
    \includegraphics[width=\linewidth]{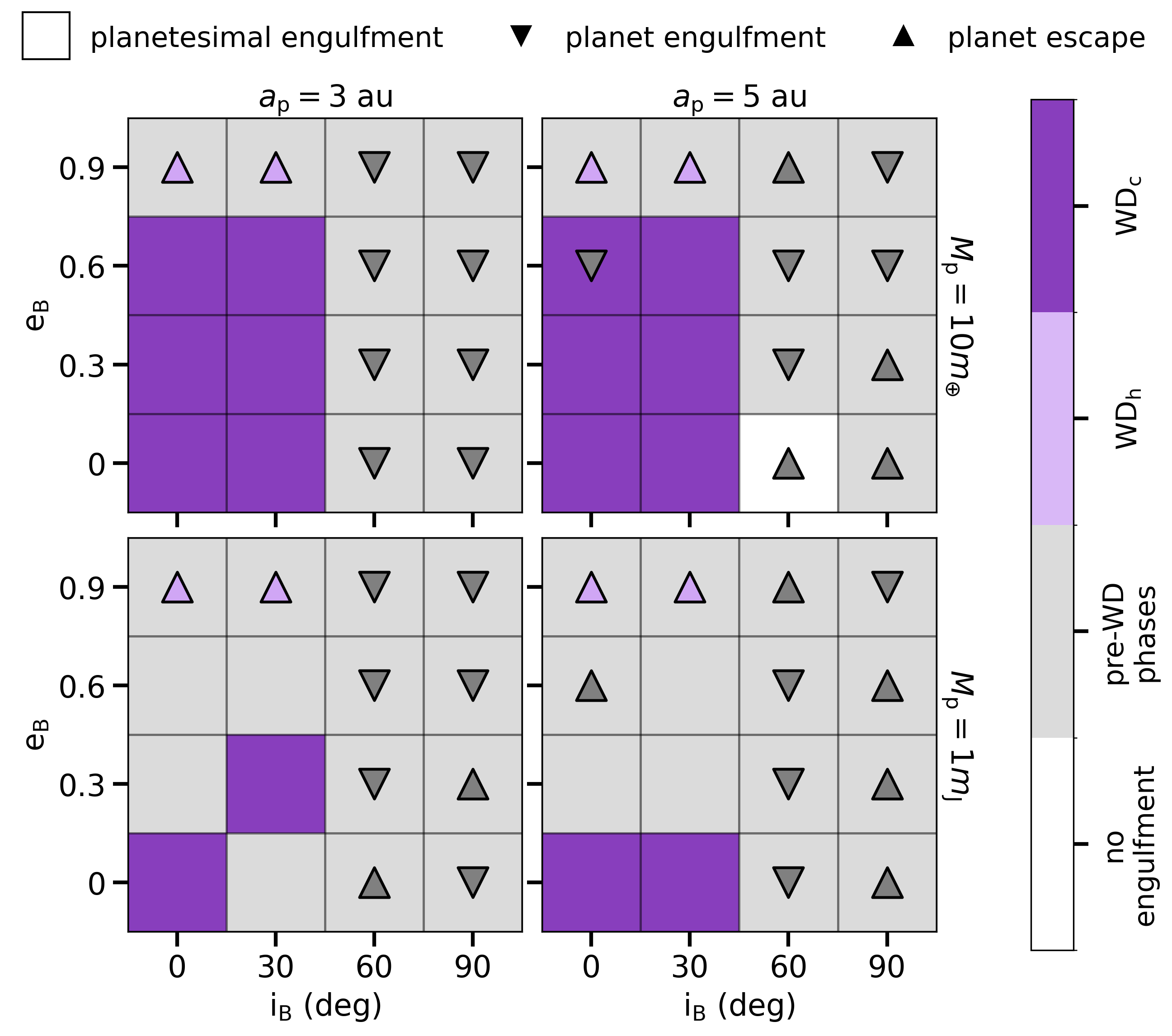}
    \caption{Simulation outcomes for the different configurations. The x- and y-axes correspond to the binary inclination $i_{\rm B}$ and eccentricity $e_{\rm B}$, respectively. Left columns correspond to planetary semi-major axes $a_{\rm p}=3$~au, and the right columns are for a planet located at $a_{\rm p}=5$~au. The top panels correspond to $m_{\rm p}=10\,m_\oplus$, while the bottom ones are for  $m_{\rm p}=1\,m_{\rm Jup}$. In all simulations, the binary is initially located at $a_{\rm B}=250$~au. The background colour indicates planetesimal engulfment, while inverted and upright triangles denote planet engulfment and planet escape, respectively. The colour indicates the evolutionary stage of the host star at the time of each event: pre-WD (gray), WD$_{\rm h}$ (light purple), and WD$_{\rm c}$ (dark purple). White cells correspond to cases in which no planetesimals are engulfed, some survive, while others are ejected from the system.} 
    \label{fig:SE_vs_Jup}
\end{figure}

The simulations start during the horizontal branch (HB) phase and follow the system through the AGB, which we refer to as the ``pre-WD phases''. During this evolution, planetesimals can be engulfed by the expanding star or ejected through dynamical interactions, becoming gravitationally unbound. We focus on the multi-body dynamical evolution during post-MS stellar evolution, when mass loss and stellar expansion can substantially reshape the system \citep{2012MNRAS.420.2990B}, assuming an initial architecture that has survived the MS.

The temporal evolution of all orbiting bodies is shown in Fig.~\ref{fig:temporal_evolution}, with pericenter distances (top) and eccentricities (bottom) as a function of stellar age. The figure shows a representative configuration with a binary (yellow) at $a_{\rm B}=250$~au, $i_{\rm B}=0^{\circ}$, and $e_{\rm B}=0.3$, and a $10\,m_\oplus$ planet (black) initially at $a_{\rm p}=3$~au, exterior to an inner disc of 25 planetesimals (gray). The colour bar at the top indicates the evolutionary stage of the host star.

The solid pink line in the top panel marks the stellar radius ($R_{\star}$), while the dashed pink line indicates the Roche radius \citep[$a_{\rm R}$,][]{1999Murray}, i.e., the minimum orbital distance before tidal disruption. 
During the pre-WD phases, the Roche limit lies within the stellar radius, and engulfment is therefore governed by $R_{\star}$ \citep{2011ApJ...737...66K,2013ApJ...777L..30A}. After WD formation, the Roche limit moves outside the stellar surface, and bodies approaching the star are expected to undergo tidal disruption instead. Since the exact accretion boundary during the WD phase remains uncertain due to the complex physics of tidal disruption and dissipation \citep{2016Veras}, we adopt $R_{\star}$ and $a_{\rm R}$ as the engulfment boundaries during the pre-WD and WD phases, respectively.

As expected from adiabatic orbital evolution driven by stellar mass loss, which predicts a maximum change in semi-major axis $M_{\rm MS}/M_{\rm WD}$ \citep{1924MNRAS..85....2J,2002Burleigh}, both the orbits of the planet and binary companion expand by factors of $\approx 2$--$3$ in all simulations. The fate of the inner planetesimals is instead determined by the competition between orbital expansion and the increasing stellar radius: for some bodies, the star overtakes their orbits before substantial orbital expansion can occur, leading to engulfment.

The efficiency of this mechanism depends on the specific architecture of the planetary system. Figure~\ref{fig:SE_vs_Jup} summarizes the outcomes for the $10\,m_\oplus$ planet (top panels) and the $1\,m_{\rm Jup}$ planet (bottom panels). The colour scheme indicates the stellar evolutionary stage at which the event occurs: gray corresponds to the pre-WD phases, light purple to the WD$_{\rm h}$ phase, and dark purple to the WD$_{\rm c}$ phase. In each cell, the background colour records the last planetesimal engulfment event. White cells correspond to cases in which no planetesimals are engulfed; the remaining planetesimals either survive or are ejected from the system. Planet-related outcomes are also shown, inverted triangles mark planet engulfments and upright triangles indicate escapes. 

Planetesimal engulfment occurs throughout the stellar evolution, from the HB to the WD$_{\rm c}$ phases. However, engulfment during the giant phases is not reflected in the atmospheric composition of the final WD. We therefore focus primarily on the dark-purple cells, which correspond to planetesimal accretion during the WD$_{\rm c}$ phase and thus represent the most relevant outcomes for observable WD pollution.

The planet acts as a dynamical link between the binary and the planetesimal disc. For low binary inclinations ($i_{\rm B}<40^\circ$), ZKL oscillations are absent and the planet remains relatively stable. In this regime, stellar mass loss and the resulting orbital expansion enhance planet--planetesimal interactions and scattering, while secular perturbations from the binary can further excite eccentricities and drive some bodies onto WD-grazing orbits \citep{2011Bonsor}.

At higher inclinations ($i_{\rm B}\ge40^\circ$), the planet enters the ZKL regime, where the binary drives coupled eccentricity--inclination oscillations. The resulting eccentricity excitation can reduce the planetary pericenter, leading to tidal interactions, engulfment, or ejection. At this binary separation, ZKL oscillations are triggered during the pre-WD stages for all planet masses, causing planetary engulfment during these phases. The planet's dynamical excitation, in turn, perturbs the planetesimal population and can drive some bodies onto star-grazing orbits. Planetesimal accretion during the WD$_{\rm c}$ phase is more frequent for lower-mass planets than for giant planets. 

The results show that the fate of the planetesimal discs is primarily determined by the binary inclination and eccentricity, as well as by the mass of the planet. These parameters control the strength and timescale of the dynamical perturbations experienced by both the planet and the planetesimal reservoir, and therefore strongly influence the efficiency and duration of planetesimal delivery to the white dwarf. In contrast, changing the planet's location from 3~au (left panels) to 5~au (right panels) has a comparatively minor impact on the overall pollution outcomes. This suggests that, within the range of orbital distances explored here, the binary configuration and planetary mass play a more important role in shaping the long-term evolution of the planetesimal reservoir than the precise location of the planet.

The planet mass also determines the lifetime of the planetesimal reservoir. For a Jupiter-mass planet, the planetesimals are efficiently destabilized and the reservoir is rapidly depleted, limiting the duration of the pollution phase, as depicted with the gray background colours in Fig.~\ref{fig:SE_vs_Jup}. In contrast, for a $10\,m_\oplus$ planet, the weaker dynamical perturbations allow a significant fraction of the planetesimal population to survive. Although some planetesimals are delivered to the star on delayed timescales during the WD$_{\rm c}$ phase, many remain in the system, providing a sustained reservoir that can continue to feed atmospheric pollution at later evolutionary stages.

Planetary survival also depends on the adopted system architecture and stellar evolution model. Differences in the adopted stellar evolution prescription may alter the competition between orbital expansion due to mass-loss and orbital decay due to tides \citep{2012Mustill}. Our results similarly suggest that the critical distance for planetary survival is not universal, but depends on the dynamical architecture of the system, particularly on $e_{\rm B}$ and $i_{\rm B}$. More generally, explaining pollution across the full range of WD cooling ages requires both a long-lived planetesimal reservoir and an efficient mechanism for sustained delivery \citep{2014MNRAS.439.3371W,2014Koester,2017Petrovich,2022MNRAS.513.4178O}. At the binary separation explored here, such long-term delivery is most efficient in the super-Earth regime, where the weaker perturbations exerted by the planet allow a significant fraction of the planetesimal reservoir to survive and remain available for later accretion. 

A more extensive exploration of the dependence on the binary configuration is presented in \citet{2026Norambuena}, where we investigate a broader range of binary separations and configurations for a Jupiter-mass planet, while also considering different stellar evolutionary mass-loss prescriptions. That study shows that the dynamical pathway to pollution also depends strongly on the binary separation. For short binary separations, both our results and those of \citet{2026Norambuena} show that pollution is only possible when the binary does not induce strong dynamical perturbations on the planetary system. In contrast, for wider binaries, pollution of the central star is primarily triggered once ZKL oscillations become active, driving planetesimals onto orbits with sufficiently short pericenter distances to eventually reach the stellar surface.

\section{Conclusions}
\label{sec:conclusion}

Our results show that binary architecture controls the dynamical excitation and delivery of planetesimals, while planetary mass determines the lifetime of the planetesimal reservoir. For a $10\,m_\oplus$ planet, a substantial fraction of the reservoir survives beyond the WD$_{\rm c}$ phase, allowing delayed delivery and long-lived atmospheric pollution, whereas a Jupiter-mass planet rapidly depletes the reservoir. The planet's location has a comparatively minor impact within the range explored here. These results highlight the need to jointly consider planetary architecture and binary dynamics when interpreting polluted white dwarfs in binary systems.

\begin{acknowledgements}
C.N. and C.A.G. acknowledge support from Agencia Nacional de Investigación y Desarrollo (ANID) through FONDECYT Iniciación 11230741 and FONDECYT Regular 1262342. 
C.C. acknowledges support from CAS-ANID (CASSACA) n$^\circ$250005.
\end{acknowledgements}


\bibliographystyle{iaulike}
\bibliography{biblio}

@ARTICLE{1910vonZeipel,
       author = {{von Zeipel}, H.},
        title = "{Sur l'application des s{\'e}ries de M. Lindstedt {\`a} l'{\'e}tude du mouvement des com{\`e}tes p{\'e}riodiques}",
      journal = {Astronomische Nachrichten},
         year = 1910,
        month = mar,
       volume = {183},
       number = {22},
        pages = {345},
          doi = {10.1002/asna.19091832202},
       adsurl = {https://ui.adsabs.harvard.edu/abs/1910AN....183..345V}
}

@ARTICLE{1924MNRAS..85....2J,
       author = {{Jeans}, J.~H.},
        title = "{Cosmogonic problems associated with a secular decrease of mass}",
      journal = {\mnras},
         year = 1924,
        month = nov,
       volume = {85},
        pages = {2},
          doi = {10.1093/mnras/85.1.2},
       adsurl = {https://ui.adsabs.harvard.edu/abs/1924MNRAS..85....2J}
}

@ARTICLE{1962Lidov,
       author = {{Lidov}, M.~L.},
        title = "{The evolution of orbits of artificial satellites of planets under the action of gravitational perturbations of external bodies}",
      journal = {\planss},
         year = 1962,
        month = oct,
       volume = {9},
       number = {10},
        pages = {719-759},
          doi = {10.1016/0032-0633(62)90129-0},
       adsurl = {https://ui.adsabs.harvard.edu/abs/1962P&SS....9..719L}
}

@ARTICLE{1962Kozai,
       author = {{Kozai}, Yoshihide},
        title = "{Secular perturbations of asteroids with high inclination and eccentricity}",
      journal = {\aj},
         year = 1962,
        month = nov,
       volume = {67},
        pages = {591-598},
          doi = {10.1086/108790},
       adsurl = {https://ui.adsabs.harvard.edu/abs/1962AJ.....67..591K}
}

@INCOLLECTION{1975Reimers,
       author = {{Reimers}, D.},
        title = "{Circumstellar envelopes and mass loss of red giant stars.}",
    booktitle = {Problems in stellar atmospheres and envelopes.},
         year = 1975,
       editor = {{Baschek}, B. and {Kegel}, W.~H. and {Traving}, G.},
        pages = {229-256},
       adsurl = {https://ui.adsabs.harvard.edu/abs/1975psae.book..229R}
}

@ARTICLE{1977Zahn,
       author = {{Zahn}, J.-P.},
        title = "{Tidal friction in close binary systems.}",
      journal = {\aap},
         year = 1977,
        month = may,
       volume = {57},
        pages = {383-394},
       adsurl = {https://ui.adsabs.harvard.edu/abs/1977A&A....57..383Z}
}

@ARTICLE{1979Mignard,
       author = {{Mignard}, F.},
        title = "{The Evolution of the Lunar Orbit Revisited. I.}",
      journal = {Moon and Planets},
         year = 1979,
        month = may,
       volume = {20},
       number = {3},
        pages = {301-315},
          doi = {10.1007/BF00907581},
       adsurl = {https://ui.adsabs.harvard.edu/abs/1979M&P....20..301M}
}

@ARTICLE{1981Hut,
       author = {{Hut}, P.},
        title = "{Tidal evolution in close binary systems.}",
      journal = {\aap},
         year = 1981,
        month = jun,
       volume = {99},
        pages = {126-140},
       adsurl = {https://ui.adsabs.harvard.edu/abs/1981A&A....99..126H}
}

@ARTICLE{1995Blocker,
       author = {{Bloecker}, T.},
        title = "{Stellar evolution of low- and intermediate-mass stars. II. Post-AGB evolution.}",
      journal = {\aap},
         year = 1995,
        month = jul,
       volume = {299},
        pages = {755},
       adsurl = {https://ui.adsabs.harvard.edu/abs/1995A&A...299..755B}
}

@ARTICLE{1995Chayer,
       author = {{Chayer}, P. and {Fontaine}, G. and {Wesemael}, F.},
        title = "{Radiative Levitation in Hot White Dwarfs: Equilibrium Theory}",
      journal = {\apjs},
         year = 1995,
        month = jul,
       volume = {99},
        pages = {189},
          doi = {10.1086/192184},
       adsurl = {https://ui.adsabs.harvard.edu/abs/1995ApJS...99..189C}
}

@BOOK{1999Murray,
       author = {{Murray}, Carl D. and {Dermott}, Stanley F.},
        title = "{Solar System Dynamics}",
         year = 1999,
          doi = {10.1017/CBO9781139174817},
       adsurl = {https://ui.adsabs.harvard.edu/abs/1999ssd..book.....M}
}

@article{2002Burleigh,
    author = {Burleigh, M. R. and Clarke, F. J. and Hodgkin, S. T.},
    title = {Imaging planets around nearby white dwarfs},
    journal = {Monthly Notices of the Royal Astronomical Society},
    volume = {331},
    number = {4},
    pages = {L41-L45},
    year = {2002},
    month = {04},
    issn = {0035-8711},
    doi = {10.1046/j.1365-8711.2002.05417.x},
    url = {https://doi.org/10.1046/j.1365-8711.2002.05417.x},
    eprint = {https://academic.oup.com/mnras/article-pdf/331/4/L41/3795814/331-4-L41.pdf},
}

@ARTICLE{2007Zuckerman,
       author = {{Zuckerman}, B. and {Koester}, D. and {Melis}, C. and {Hansen}, Brad M. and {Jura}, M.},
        title = "{The Chemical Composition of an Extrasolar Minor Planet}",
      journal = {\apj},
         year = 2007,
        month = dec,
       volume = {671},
       number = {1},
        pages = {872-877},
          doi = {10.1086/522223},
archivePrefix = {arXiv},
       eprint = {0708.0198},
 primaryClass = {astro-ph},
       adsurl = {https://ui.adsabs.harvard.edu/abs/2007ApJ...671..872Z}
}

@ARTICLE{2009Koester,
       author = {{Koester}, D.},
        title = "{Accretion and diffusion in white dwarfs. New diffusion timescales and applications to GD 362 and G 29-38}",
      journal = {\aap},
         year = 2009,
        month = may,
       volume = {498},
       number = {2},
        pages = {517-525},
          doi = {10.1051/0004-6361/200811468},
archivePrefix = {arXiv},
       eprint = {0903.1499},
 primaryClass = {astro-ph.SR},
       adsurl = {https://ui.adsabs.harvard.edu/abs/2009A&A...498..517K}
}

@ARTICLE{2011Bonsor,
       author = {{Bonsor}, A. and {Mustill}, A.~J. and {Wyatt}, M.~C.},
        title = "{Dynamical effects of stellar mass-loss on a Kuiper-like belt}",
      journal = {\mnras},
         year = 2011,
        month = jun,
       volume = {414},
       number = {2},
        pages = {930-939},
          doi = {10.1111/j.1365-2966.2011.18524.x},
archivePrefix = {arXiv},
       eprint = {1102.3185},
 primaryClass = {astro-ph.EP},
       adsurl = {https://ui.adsabs.harvard.edu/abs/2011MNRAS.414..930B}
}

@ARTICLE{2011ApJ...737...66K,
       author = {{Kunitomo}, M. and {Ikoma}, M. and {Sato}, B. and {Katsuta}, Y. and {Ida}, S.},
        title = "{Planet Engulfment by \raisebox{-0.5ex}\textasciitilde1.5-3 M $_{sun}$ Red Giants}",
      journal = {\apj},
         year = 2011,
        month = aug,
       volume = {737},
       number = {2},
          eid = {66},
        pages = {66},
          doi = {10.1088/0004-637X/737/2/66},
archivePrefix = {arXiv},
       eprint = {1106.2251},
 primaryClass = {astro-ph.EP},
       adsurl = {https://ui.adsabs.harvard.edu/abs/2011ApJ...737...66K}
}

@ARTICLE{Paxton2011,
  author = {{Paxton}, B. and {Bildsten}, L. and {Dotter}, A. and {Herwig}, F. and {Lesaffre}, P. and {Timmes}, F.},
  title = {{Modules for Experiments in Stellar Astrophysics (MESA)}},
  journal = {\apjs},
  archivePrefix = {arXiv},
  eprint = {1009.1622},
  primaryClass = {astro-ph.SR},
  year = {2011},
  month = {jan},
  volume = {192},
  eid = {3},
  pages = {3},
  doi = {10.1088/0067-0049/192/1/3},
  adsurl = {https://ui.adsabs.harvard.edu/abs/2011ApJS..192....3P},
}

@ARTICLE{2012MNRAS.420.2990B,
       author = {{Bonsor}, A. and {Wyatt}, M.~C.},
        title = "{The scattering of small bodies in planetary systems: constraints on the possible orbits of cometary material}",
      journal = {\mnras},
         year = 2012,
        month = mar,
       volume = {420},
       number = {4},
        pages = {2990-3002},
          doi = {10.1111/j.1365-2966.2011.20156.x},
archivePrefix = {arXiv},
       eprint = {1111.1858},
 primaryClass = {astro-ph.EP},
       adsurl = {https://ui.adsabs.harvard.edu/abs/2012MNRAS.420.2990B}
}

@ARTICLE{2012Mustill,
       author = {{Mustill}, Alexander J. and {Villaver}, Eva},
        title = "{Foretellings of Ragnar{\"o}k: World-engulfing Asymptotic Giants and the Inheritance of White Dwarfs}",
      journal = {\apj},
         year = 2012,
        month = dec,
       volume = {761},
       number = {2},
          eid = {121},
        pages = {121},
          doi = {10.1088/0004-637X/761/2/121},
archivePrefix = {arXiv},
       eprint = {1210.0328},
 primaryClass = {astro-ph.EP},
       adsurl = {https://ui.adsabs.harvard.edu/abs/2012ApJ...761..121M}
}

@ARTICLE{2012ReinLiu,
       author = {{Rein}, H. and {Liu}, S.-F.},
        title = "{REBOUND: an open-source multi-purpose N-body code for collisional dynamics}",
      journal = {\aap},
         year = 2012,
        month = jan,
       volume = {537},
          eid = {A128},
        pages = {A128},
          doi = {10.1051/0004-6361/201118085},
archivePrefix = {arXiv},
       eprint = {1110.4876},
 primaryClass = {astro-ph.EP},
       adsurl = {https://ui.adsabs.harvard.edu/abs/2012A&A...537A.128R}
}

@ARTICLE{2013ApJ...777L..30A,
       author = {{Adams}, Fred C. and {Bloch}, Anthony M.},
        title = "{Evolution of Planetary Orbits with Stellar Mass Loss and Tidal Dissipation}",
      journal = {\apjl},
         year = 2013,
        month = nov,
       volume = {777},
       number = {2},
          eid = {L30},
        pages = {L30},
          doi = {10.1088/2041-8205/777/2/L30},
archivePrefix = {arXiv},
       eprint = {1310.2577},
 primaryClass = {astro-ph.SR},
       adsurl = {https://ui.adsabs.harvard.edu/abs/2013ApJ...777L..30A}
}

@ARTICLE{2013Farihi,
       author = {{Farihi}, J. and {G{\"a}nsicke}, B.~T. and {Koester}, D.},
        title = "{Evidence for Water in the Rocky Debris of a Disrupted Extrasolar Minor Planet}",
      journal = {Science},
         year = 2013,
        month = oct,
       volume = {342},
       number = {6155},
        pages = {218-220},
          doi = {10.1126/science.1239447},
archivePrefix = {arXiv},
       eprint = {1310.3269},
 primaryClass = {astro-ph.EP},
       adsurl = {https://ui.adsabs.harvard.edu/abs/2013Sci...342..218F}
}

@ARTICLE{2014Jura,
       author = {{Jura}, M. and {Young}, E.~D.},
        title = "{Extrasolar Cosmochemistry}",
      journal = {Annual Review of Earth and Planetary Sciences},
         year = 2014,
        month = may,
       volume = {42},
       number = {1},
        pages = {45-67},
          doi = {10.1146/annurev-earth-060313-054740},
       adsurl = {https://ui.adsabs.harvard.edu/abs/2014AREPS..42...45J}
}

@ARTICLE{2014Koester,
       author = {{Koester}, D. and {G{\"a}nsicke}, B.~T. and {Farihi}, J.},
        title = "{The frequency of planetary debris around young white dwarfs}",
      journal = {\aap},
         year = 2014,
        month = jun,
       volume = {566},
          eid = {A34},
        pages = {A34},
          doi = {10.1051/0004-6361/201423691},
archivePrefix = {arXiv},
       eprint = {1404.2617},
 primaryClass = {astro-ph.SR},
       adsurl = {https://ui.adsabs.harvard.edu/abs/2014A&A...566A..34K}
}

@ARTICLE{2014MNRAS.439.3371W,
       author = {{Wyatt}, M.~C. and {Farihi}, J. and {Pringle}, J.~E. and {Bonsor}, A.},
        title = "{Stochastic accretion of planetesimals on to white dwarfs: constraints on the mass distribution of accreted material from atmospheric pollution}",
      journal = {\mnras},
         year = 2014,
        month = apr,
       volume = {439},
       number = {4},
        pages = {3371-3391},
          doi = {10.1093/mnras/stu183},
archivePrefix = {arXiv},
       eprint = {1401.6173},
 primaryClass = {astro-ph.EP},
       adsurl = {https://ui.adsabs.harvard.edu/abs/2014MNRAS.439.3371W}
}

@ARTICLE{2014Xu,
       author = {{Xu}, S. and {Jura}, M. and {Koester}, D. and {Klein}, B. and {Zuckerman}, B.},
        title = "{Elemental Compositions of Two Extrasolar Rocky Planetesimals}",
      journal = {\apj},
         year = 2014,
        month = mar,
       volume = {783},
       number = {2},
          eid = {79},
        pages = {79},
          doi = {10.1088/0004-637X/783/2/79},
archivePrefix = {arXiv},
       eprint = {1401.4252},
 primaryClass = {astro-ph.EP},
       adsurl = {https://ui.adsabs.harvard.edu/abs/2014ApJ...783...79X}
}

@ARTICLE{2015ReinSpiegel,
       author = {{Rein}, Hanno and {Spiegel}, David S.},
        title = "{IAS15: a fast, adaptive, high-order integrator for gravitational dynamics, accurate to machine precision over a billion orbits}",
      journal = {\mnras},
         year = 2015,
        month = jan,
       volume = {446},
       number = {2},
        pages = {1424-1437},
          doi = {10.1093/mnras/stu2164},
archivePrefix = {arXiv},
       eprint = {1409.4779},
 primaryClass = {astro-ph.EP},
       adsurl = {https://ui.adsabs.harvard.edu/abs/2015MNRAS.446.1424R}
}

@ARTICLE{2015ReinTamayo,
       author = {{Rein}, Hanno and {Tamayo}, Daniel},
        title = "{WHFAST: a fast and unbiased implementation of a symplectic Wisdom-Holman integrator for long-term gravitational simulations}",
      journal = {\mnras},
         year = 2015,
        month = sep,
       volume = {452},
       number = {1},
        pages = {376-388},
          doi = {10.1093/mnras/stv1257},
archivePrefix = {arXiv},
       eprint = {1506.01084},
 primaryClass = {astro-ph.EP},
       adsurl = {https://ui.adsabs.harvard.edu/abs/2015MNRAS.452..376R}
}

@ARTICLE{2015Vanderburg,
       author = {{Vanderburg}, Andrew and {Johnson}, John Asher and {Rappaport}, Saul and {Bieryla}, Allyson and {Irwin}, Jonathan and {Lewis}, John Arban and {Kipping}, David and {Brown}, Warren R. and {Dufour}, Patrick and {Ciardi}, David R. and {Angus}, Ruth and {Schaefer}, Laura and {Latham}, David W. and {Charbonneau}, David and {Beichman}, Charles and {Eastman}, Jason and {McCrady}, Nate and {Wittenmyer}, Robert A. and {Wright}, Jason T.},
        title = "{A disintegrating minor planet transiting a white dwarf}",
      journal = {\nat},
         year = 2015,
        month = oct,
       volume = {526},
       number = {7574},
        pages = {546-549},
          doi = {10.1038/nature15527},
archivePrefix = {arXiv},
       eprint = {1510.06387},
 primaryClass = {astro-ph.EP},
       adsurl = {https://ui.adsabs.harvard.edu/abs/2015Natur.526..546V}
}

@ARTICLE{2016Farihi,
       author = {{Farihi}, J.},
        title = "{Circumstellar debris and pollution at white dwarf stars}",
      journal = {\nar},
         year = 2016,
        month = apr,
       volume = {71},
        pages = {9-34},
          doi = {10.1016/j.newar.2016.03.001},
archivePrefix = {arXiv},
       eprint = {1604.03092},
 primaryClass = {astro-ph.EP},
       adsurl = {https://ui.adsabs.harvard.edu/abs/2016NewAR..71....9F}
}

@ARTICLE{2016Veras,
       author = {{Veras}, Dimitri},
        title = "{Post-main-sequence planetary system evolution}",
      journal = {Royal Society Open Science},
         year = 2016,
        month = feb,
       volume = {3},
          eid = {150571},
        pages = {150571},
          doi = {10.1098/rsos.150571},
archivePrefix = {arXiv},
       eprint = {1601.05419},
 primaryClass = {astro-ph.EP},
       adsurl = {https://ui.adsabs.harvard.edu/abs/2016RSOS....350571V}
}

@ARTICLE{2017Hollands,
       author = {{Hollands}, M.~A. and {Koester}, D. and {Alekseev}, V. and {Herbert}, E.~L. and {G{\"a}nsicke}, B.~T.},
        title = "{Cool DZ white dwarfs - I. Identification and spectral analysis}",
      journal = {\mnras},
         year = 2017,
        month = jun,
       volume = {467},
       number = {4},
        pages = {4970-5000},
          doi = {10.1093/mnras/stx250},
archivePrefix = {arXiv},
       eprint = {1701.07827},
 primaryClass = {astro-ph.SR},
       adsurl = {https://ui.adsabs.harvard.edu/abs/2017MNRAS.467.4970H}
}

@ARTICLE{2017Petrovich,
       author = {{Petrovich}, Cristobal and {Mu{\~n}oz}, Diego J.},
        title = "{Planetary Engulfment as a Trigger for White Dwarf Pollution}",
      journal = {\apj},
         year = 2017,
        month = jan,
       volume = {834},
       number = {2},
          eid = {116},
        pages = {116},
          doi = {10.3847/1538-4357/834/2/116},
archivePrefix = {arXiv},
       eprint = {1607.04891},
 primaryClass = {astro-ph.EP},
       adsurl = {https://ui.adsabs.harvard.edu/abs/2017ApJ...834..116P}
}

@ARTICLE{2017Stephan,
       author = {{Stephan}, Alexander P. and {Naoz}, Smadar and {Zuckerman}, B.},
        title = "{Throwing Icebergs at White Dwarfs}",
      journal = {\apjl},
         year = 2017,
        month = aug,
       volume = {844},
       number = {2},
          eid = {L16},
        pages = {L16},
          doi = {10.3847/2041-8213/aa7cf3},
archivePrefix = {arXiv},
       eprint = {1704.08701},
 primaryClass = {astro-ph.SR},
       adsurl = {https://ui.adsabs.harvard.edu/abs/2017ApJ...844L..16S}
}

@ARTICLE{2017Veras,
       author = {{Veras}, Dimitri and {Georgakarakos}, Nikolaos and {Dobbs-Dixon}, Ian and {G{\"a}nsicke}, Boris T.},
        title = "{Binary star influence on post-main-sequence multi-planet stability}",
      journal = {\mnras},
         year = 2017,
        month = feb,
       volume = {465},
       number = {2},
        pages = {2053-2059},
          doi = {10.1093/mnras/stw2699},
archivePrefix = {arXiv},
       eprint = {1610.05307},
 primaryClass = {astro-ph.EP},
       adsurl = {https://ui.adsabs.harvard.edu/abs/2017MNRAS.465.2053V}
}

@ARTICLE{2018Mustill,
       author = {{Mustill}, Alexander J. and {Villaver}, Eva and {Veras}, Dimitri and {G{\"a}nsicke}, Boris T. and {Bonsor}, Amy},
        title = "{Unstable low-mass planetary systems as drivers of white dwarf pollution}",
      journal = {\mnras},
         year = 2018,
        month = may,
       volume = {476},
       number = {3},
        pages = {3939-3955},
          doi = {10.1093/mnras/sty446},
archivePrefix = {arXiv},
       eprint = {1711.02940},
 primaryClass = {astro-ph.EP},
       adsurl = {https://ui.adsabs.harvard.edu/abs/2018MNRAS.476.3939M}
}

@ARTICLE{2018Veras,
       author = {{Veras}, Dimitri and {Xu}, Siyi and {Rebassa-Mansergas}, Alberto},
        title = "{The critical binary star separation for a planetary system origin of white dwarf pollution}",
      journal = {\mnras},
         year = 2018,
        month = jan,
       volume = {473},
       number = {3},
        pages = {2871-2880},
          doi = {10.1093/mnras/stx2141},
archivePrefix = {arXiv},
       eprint = {1708.05391},
 primaryClass = {astro-ph.EP},
       adsurl = {https://ui.adsabs.harvard.edu/abs/2018MNRAS.473.2871V}
}

@ARTICLE{2019Wilson,
       author = {{Wilson}, Thomas G. and {Farihi}, Jay and {G{\"a}nsicke}, Boris T. and {Swan}, Andrew},
        title = "{The unbiased frequency of planetary signatures around single and binary white dwarfs using Spitzer and Hubble}",
      journal = {\mnras},
         year = 2019,
        month = jul,
       volume = {487},
       number = {1},
        pages = {133-146},
          doi = {10.1093/mnras/stz1050},
archivePrefix = {arXiv},
       eprint = {1904.05891},
 primaryClass = {astro-ph.EP},
       adsurl = {https://ui.adsabs.harvard.edu/abs/2019MNRAS.487..133W}
}

@ARTICLE{2020Ronco,
       author = {{Ronco}, Mar{\'\i}a Paula and {Schreiber}, Matthias R. and {Giuppone}, Cristian A. and {Veras}, Dimitri and {Cuadra}, Jorge and {Guilera}, Octavio M.},
        title = "{How Jupiters Save or Destroy Inner Neptunes around Evolved Stars}",
      journal = {\apjl},
         year = 2020,
        month = jul,
       volume = {898},
       number = {1},
          eid = {L23},
        pages = {L23},
          doi = {10.3847/2041-8213/aba35f},
archivePrefix = {arXiv},
       eprint = {2007.04337},
 primaryClass = {astro-ph.EP},
       adsurl = {https://ui.adsabs.harvard.edu/abs/2020ApJ...898L..23R}
}

@ARTICLE{2020Tamayo,
       author = {{Tamayo}, Daniel and {Rein}, Hanno and {Shi}, Pengshuai and {Hernandez}, David M.},
        title = "{REBOUNDx: a library for adding conservative and dissipative forces to otherwise symplectic N-body integrations}",
      journal = {\mnras},
         year = 2020,
        month = jan,
       volume = {491},
       number = {2},
        pages = {2885-2901},
          doi = {10.1093/mnras/stz2870},
archivePrefix = {arXiv},
       eprint = {1908.05634},
 primaryClass = {astro-ph.EP},
       adsurl = {https://ui.adsabs.harvard.edu/abs/2020MNRAS.491.2885T}
}

@ARTICLE{2020Vanderbosch,
       author = {{Vanderbosch}, Z. and {Hermes}, J.~J. and {Dennihy}, E. and {Dunlap}, B.~H. and {Izquierdo}, P. and {Tremblay}, P.-E. and {Cho}, P.~B. and {G{\"a}nsicke}, B.~T. and {Toloza}, O. and {Bell}, K.~J. and {Montgomery}, M.~H. and {Winget}, D.~E.},
        title = "{A White Dwarf with Transiting Circumstellar Material Far outside the Roche Limit}",
      journal = {\apj},
         year = 2020,
        month = jul,
       volume = {897},
       number = {2},
          eid = {171},
        pages = {171},
          doi = {10.3847/1538-4357/ab9649},
archivePrefix = {arXiv},
       eprint = {1908.09839},
 primaryClass = {astro-ph.SR},
       adsurl = {https://ui.adsabs.harvard.edu/abs/2020ApJ...897..171V}
}

@ARTICLE{2022Baronett,
       author = {{Baronett}, Stanley A. and {Ferich}, Noah and {Tamayo}, Daniel and {Steffen}, Jason H.},
        title = "{Stellar evolution and tidal dissipation in REBOUNDx}",
      journal = {\mnras},
         year = 2022,
        month = mar,
       volume = {510},
       number = {4},
        pages = {6001-6009},
          doi = {10.1093/mnras/stac043},
archivePrefix = {arXiv},
       eprint = {2101.12277},
 primaryClass = {astro-ph.SR},
       adsurl = {https://ui.adsabs.harvard.edu/abs/2022MNRAS.510.6001B}
}

\end{document}